\documentclass[11pt,preprint]{aastex}

\usepackage{natbib,graphicx,epsfig}

\newcommand{\psim}{\lower.5ex\hbox{$\; \buildrel \propto \over\sim \;$}}

\shorttitle{Gamma Rays from UHECRs in Cygnus A}
\shortauthors{Atoyan \& Dermer}

\begin{document}

\title{Gamma Rays from Ultra-High Energy Cosmic Rays in Cygnus A}

\author{Armen Atoyan\altaffilmark{1} and Charles D. Dermer\altaffilmark{2}}
\altaffiltext{1}{Department of Mathematics, Concordia University,
1455 de Maisonneuve Blvd.\ West,\\ Montreal, Quebec H3G 1M8, Canada;
~atoyan@mathstat.concordia.ca}

\altaffiltext{2}{Space Science Division,
Code 7653, Naval Research Laboratory,\\ Washington, DC 20375-5352, USA;~
charles.dermer@nrl.navy.mil}

\begin{abstract}
Ultra-high energy cosmic rays (UHECRs) accelerated in the jets of active galactic
nuclei can accumulate in high magnetic field, $\sim 100$ kpc-scale regions
surrounding powerful radio galaxies. Photohadronic processes { involving 
UHECRs and photons of the extragalactic background light}
make ultra-relativistic electrons and positrons that initiate 
electromagnetic cascades,
leading to the production of a $\gamma$-ray synchrotron halo.
We calculate the halo emission in the case of Cygnus A and show that
it should be detectable with the { Fermi Gamma ray Space Telescope}
and possibly detectable with ground-based
$\gamma$-ray telescopes if radio galaxies are the sources of UHECRs.
\end{abstract}

\keywords{Galaxies: individual (Cygnus A) --- cosmic rays: theory ---
galaxies: active --- galaxies: jets --- radiation processes: nonthermal}



\section{Introduction}

Active galactic nuclei (AGN) and gamma-ray bursts are considered
as two of the most plausible classes of astrophysical accelerators of
extragalactic ultra-high energy cosmic rays \citep[UHECRs; see, e.g.,][]{hh02}.
The recent report of the \citet{Auger07} about clustering of the arrival directions
of UHECRs with energies $E \gtrsim 6\times 10^{19}$ eV within $\approx 3^\circ$
of the directions to AGN at distances $d\lesssim 75\, \rm Mpc$
strongly suggests that effective production of cosmic rays
with energies  $E\sim 10^{20}\,\rm eV$ takes place in at least one of these
source classes. Because of photohadronic GZK interactions of protons or ions
with the CMB radiation,  
the study of super-GZK UHECRs from sources at $d\gtrsim 100\,\rm Mpc$ becomes impossible
with cosmic-ray detectors like the Pierre Auger Observatory \citep{hmr06}. 
Powerful AGN, as well as GRBs, are mostly located at larger distances.

{ Relativistic beams of energy from the central nuclei of AGN are thought to 
power the multi-kpc scale radio lobes of powerful galaxies and form 
an extended cavity \citep{sch74}.
Acceleration of UHECRs in the compact inner jets of the radio galaxy 
on the sub-parsec scale, followed by 
production of collimated beams of ultra-high energy neutrons and 
gamma-rays, provides a specific mechanism to transport energy to 
the radio lobes and cavity \citep{ad03}.} 
When the neutron-decay UHECR protons interact with the extragalactic background light (EBL), 
which is dominated by the CMB radiation,
ultra-relativistic electrons (including positrons) and $\gamma$ rays with
$E\gtrsim 10^{18}\,\rm eV$ are produced. Such secondaries can initiate
pair-photon cascades to form large multi-Mpc scale halos of GeV/TeV radiation due
to Compton and synchrotron processes \citep{acv94,aha02,ias05}.

In this Letter, we predict that synchrotron GeV fluxes from UHECR AGN sources are 
detectable with the Fermi Gamma ray Space Telescope (FGST; 
formerly the Gamma ray Large 
Area Space Telescope, GLAST) if UHECRs are captured in the 
vicinity of radio galaxies for sufficiently long times.
Magnetic fields at the
$\gtrsim \mu$G level in the
kpc -- Mpc vicinity from the AGN core  are required to
isotropize UHECRs accelerated by jets of radio galaxies.  Indeed, for protons with
energy $E\equiv 10^{20} E_{20} \,\rm eV$, the mean magnetic field $B$ required
to provide gyroradii smaller than size $r$ is
$B\gtrsim 10^{-4} E_{20} r_{\,\rm kpc}^{-1}\,\rm G$, where
$r_{\rm kpc} =r/1\,\rm kpc$ is the spatial scale where the isotropization occurs.

Here we consider the specific case of the powerful radio galaxy Cygnus A,
{ where the mean magnetic field in the surrounding cavity could reach 10 -- 100 $\mu$G 
at 100 kpc scales}. Its properties are considered in Section 2, and calculations are presented in Section 3. 
We summarize in Section 4.

\section{Model Parameters}

Magnetic fields $B\gtrsim 1\, \mu G$ could be present at $\lesssim 1\,$Mpc scales
in clusters of galaxies \citep{kim90,fer95,fer08}. Even higher magnetic
fields, $B\sim 10^{-4}\,$G, could be present in the $r\lesssim 100\,$kpc
vicinity of cD galaxies near the center of galaxy clusters like the powerful radio
galaxy Cyg A  \citep{wil06}. In such magnetic fields, synchrotron radiation of
electrons with $E\sim 10^{18} \,\rm eV$
produced in photomeson interactions is in the TeV domain. Furthermore,
synchrotron radiation of lower energy electrons produced in
$p+\gamma\rightarrow p+ e^+ + e^-$  interactions by UHECRs extends to the GeV domain and could become
detectable with the Fermi Telescope. 
Here we consider the detectability of
these fluxes from the powerful and well-studied radio galaxy Cyg A.

\subsection {Power of the Cosmic Ray Accelerator}

Observations of Cygnus A ($z=0.056$, luminosity distance $\simeq 240 \,\rm Mpc$)
by the Chandra X-ray Observatory show that the central
$\sim 60~{\rm kpc} \times 120$ kpc size luminous prolate spheroidal region of Cygnus A, called a ``cavity," can be
understood as a shock expanding into the accretion cooling-flow gas \citep{wil06}.
The kinetic power of expansion of this X-ray cavity deduced by \citet{wil06} is
$L_{\rm exp} \approx 1.2\times 10^{46}$ erg s$^{-1}$. This is much larger than the total
radio luminosity $L_{\rm lobes}\approx 10^{45}$ erg s$^{-1}$
of the two bright radio lobes of Cygnus A \citep{pdc84}.
The value of $L_{\rm exp}$ should be considered as a {\it lower}
limit to the overall power $L_{CR}$ of cosmic rays
injected into the cavity, if the cosmic-ray power is assumed to drive the expansion of the cavity.
A few times larger power, $L_{\rm CR}\sim 4\times 10^{46}$ ergs s$^{-1}$
therefore seems a reasonable assumption.

This value is in a good agreement with the ``neutral beam'' model 
\citep{ad01,ad03},
which explains the collimated relativistic X-ray jets that remain straight up to
$\simeq 1$ Mpc distances in sources such as Pictor A \citep{wys01}
as the result of energy transport by beams of UHE neutrons and gamma-rays.
{ These linear X-ray features, surrounded by a broader and less collimated
radio structure in Pictor A, and coincident with 
narrow radio structures exhibiting bends and deflections 
in Cyg A 
\citep{car96,sb08}, 
terminate in X-ray hot spots. 
Because of the large inclination angle, the X-ray jets in Cyg A cannot 
be detected. However detection of bright X-ray hot spots 
located at $\simeq 60 \,\rm kpc$ distances on opposite sides
of the nucleus \citep{wil06} strongly implies production of  
collimated X-ray jets also in Cyg A.
}

Detailed calculations in the framework of this model show that 
neutral beams of ultrarelativistic neutrons and $\gamma$ rays,
produced by the compact relativistic jets in the central sub-parsec scale
environment of FRII galaxies, can take a few percent of the total power
{ of this inner jet. That energy is
then deposited, after $\beta$-decay of neutrons
$n\rightarrow p+ e +\nu_e$, on distance scales
$l_d \simeq E_{20}\,\rm Mpc$, and also via photopair production of UHE
gamma-rays through the process $\gamma+\gamma \rightarrow e^- +e^+$.
These secondary charged relativistic particles initially form a beam 
in the same direction as the jet,
and can effectively interact with and transfer momentum and energy to the ambient 
magnetized medium, and thus can be the basic energy 
reservoir for the jet at large distances from Cyg A.}
The total radio luminosity $\simeq 9\times 10^{44}\, \rm erg \, s^{-1}$
detected  from the lobes of Cyg A \citep{cb96,wil06} imposes the minimum
power requirement for the  beam. Given the $\sim (2-3)\%$ efficiency 
of neutral beam production, the acceleration power of the CRs in the
relativistic inner jet must be 
$L_{\rm CR}\gtrsim (3-5)\times 10^{46} \,\rm erg/s$.
{ This power is then released in CRs when the inner jet 
decelerates to subrelativistic speeds in the dense medium at kpc 
scales.}
This scenario is in agreement with the assumption that the expansion 
of the cavity against
the cooling flow observed \citep{Smith02}
at distances $\lesssim 70$ -- 100 kpc is powered by cosmic-ray pressure.

{ Protons are accelerated in the inner jet of Cyg A 
to a maximum energy $E_{max}\approx 10^{20}$ eV, consistent with size scales 
and magnetic fields inferred from a synchrotron model of blazars corresponding
to Cyg A if observed along its jet. 
This value of $E_{max}$ is in accord with
the spatial extent of the X-ray hot spot where the large-scale jet terminates
due to the $\beta$-decay of the remaining high-energy neutrons in the beam.
At this distance, the injection power from the decaying neutrons is
balanced by the ram pressure of the
external medium. Thus, for $l_d \sim 100$ kpc-long jets originating from
neutral beams,} the inner jets must accelerate protons to
$E\gtrsim 10^{19}$eV. In our model, we assume that acceleration of UHECRs
to $E\sim 10^{20}$ eV occurs.

\subsection {Magnetic Field Strength}

{ The lower limit to the equipartition magnetic field in the 
radio lobes of Cyg A inferred from radio observations is $\approx 60\,\rm \mu G$ \citep{car91}.
The equipartition magnetic field in the cavity derived from 
X-ray observations is $B\simeq 120\,\rm \mu G$ \citep{wil06}.}
Moreover, the field outside the cavity could also be very high
{ given the large thermal electron energy density inferred from the 
X-ray emission at $\lesssim 8^\prime $ or in the
$\lesssim 500\,$kpc
environment around Cyg A.}
The pressure of hot thermal gas decreases from $2.3\times 10^{-10}$ erg cm$^{-3}$ in the regions near
the cavity to $\lesssim 10^{-11}$ erg cm$^{-3}$ at distances $\gtrsim 500$
kpc from the center of Cyg A \citep{Smith02}.
The corresponding equipartition magnetic fields would then vary from
$\sim 80\,\rm \mu G$ to $\sim 15\,\rm \mu G$ at the periphery of this region.

\subsection{Injection Age}

The duration of injection of UHECRs, i.e.,
the jet injection age, represents one of the important parameters of the model.
\citet{wil06} find that the expansion age of the cavity,
determined from the speed of the shock deduced from the analysis of X-ray data,
is $t\sim 3\times 10^7\,$yr. The injection age can be larger than this dynamical
age because the derived value neglects 
the magnetic-field pressure of the intracluster medium
(ICM) upstream of the shock, as recognized 
by \citet{wil06}.
Also, this age estimate neglects the infall velocity of 
the cooling flow.
For the total mass $M \approx 2\times 10^{13}M_\odot$ enclosed at 
$r\leq 50\,\rm kpc$ distances \citep{Smith02}, the virial speed 
$v_{\rm vir} \approx c/\sqrt{r/r_{\rm S}} \cong 2000 \,\rm km/s$, 
{  where $r_{\rm S} = 2GM/c^2 \approx 6\times 10^{18}\,$cm  
is, formally, the Schwarzschild radius for the total mass $M$. 
The value of $v_{\rm vir}$ gives a measure of the accretion/cooling flow 
velocities at radius $r$, and is comparable to the 
average $\sim 1500 \,\rm km/s$ speed of the shock derived by \citet{wil06}
in the rest frame of the fluid upstream of the shock.}
%
These factors can significantly decrease the speed
of expansion of the cavity in the stationary frame, so that the real 
injection age of the cavity can be significantly larger than the age inferred 
by \citet{wil06}.

We now estimate the age of activity of the black-hole jet from energetics
arguments. The inferred jet power, $L_{jet} \simeq 4\times 10^{46} \,\rm erg/s$,
corresponds to  $\simeq 12\%$ of the Eddington luminosity for
a black-hole mass $M_{\rm BH} \simeq 2.5 \times 10^{9} M_{\odot}$ in Cygnus A
\citep{tud03}.
To produce such power, the black hole should accrete mass at the rate
$\dot{M}=L_{jet}/\eta c^2= (7/\eta_{-1})  M_\odot \,\rm yr^{-1}$ with an efficiency
$\eta = 10^{-1} \eta_{-1}$, 
{ with $\eta_{-1} \approx 1$}.
The age of the central black hole {  
is estimated by its growth time
$t\simeq M_{\rm BH}/\dot{M} =  3.6\times 10^{8} \eta_{-1} \,\rm yr$,
giving an upper estimate for the jet's age as the jet 
might be active for only a fraction of the BH growth phase. 
These estimates are in accord with the characteristic jet age $t\sim 10^8$ yr 
inferred from a model for ``cocoon'' 
(or cavity) dynamics by \citet{bc89}.}

\subsection{Cosmic Ray Diffusive Confinement Time}

The maximum confinement timescale $t_{conf}$ of UHECRs in a source of
size $r$ is given in the Bohm diffusion approximation by
$t_{conf} \cong r^2/2D_{\rm B}$, where the Bohm diffusion
coefficient $D_{\rm B} = c r_{\rm L}/3$,
{ and $r_L \cong E/QB$ is the Larmor radius of a particle with charge $Q$}. 
From this expression we obtain
the UHECR proton confinement time
\begin{equation}
t_{conf} \cong 0.95\times 10^7 \; {(r/100~{\rm kpc})^2 }  E_{20}^{-1}\;(B/20~\mu {\rm G})\; {\rm yr}\;.
\label{tesc}
\end{equation}
For $r\sim 50$ kpc and $B \cong 120~\mu$G, $t_{conf} \cong 15$ Myr for $\approx 10^{20}$ eV
protons.

\begin{figure}[t]
\center
\includegraphics[width= 8.0cm]{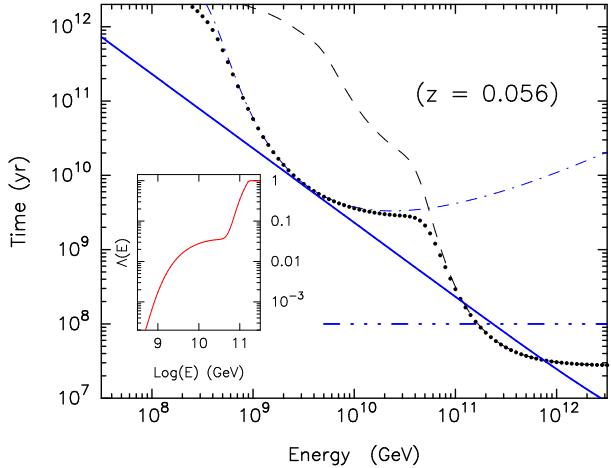}
\caption{Comparison of timescales for energy losses through 
photomeson (dashed curve)
and photopair (dot-dashed curve) production,
confinement (solid line) with $B = 20~\mu$G and $r = 500$ kpc,
and  injection with $t_{age} = 10^8$ yrs
(triple-dot-dashed line). The inset shows
the fraction $\Lambda(E)$
of energy extracted by photohadronic processes from UHECRs with
energy $E$.}
\vskip0.2in
\label{f1}
\end{figure}

Fig.\ \ref{f1} compares timescales $t_{loss}$ for energy losses due to photomeson and
photopair production
with values of $t_{conf}$ for characteristic parameters $r = 500$ kpc
and $B = 20~\mu$G in the region surrounding Cyg A.
 Also shown is the value of $t_{age} = 10^8$ yrs for the assumed injection age.
The fraction of energy of accelerated protons
that could be extracted is given by $\Lambda(E) = \min(t_{conf},t_{loss},t_{age})/t_{loss}$.
This figure shows that $\gtrsim 3$\% of the total energy of UHECRs with
$E \gtrsim 3\times 10^{18}$ eV
can be extracted through photohadronic processes. Because of the increased
confinement time and the much
lower production threshold for photopair than photomeson processes, the photopair process
can make a comparable or dominant contribution to the electromagnetic channel compared to
photomeson processes.

\section{Results}

\begin{figure}[t]
\center
\includegraphics[width= 8.0cm]{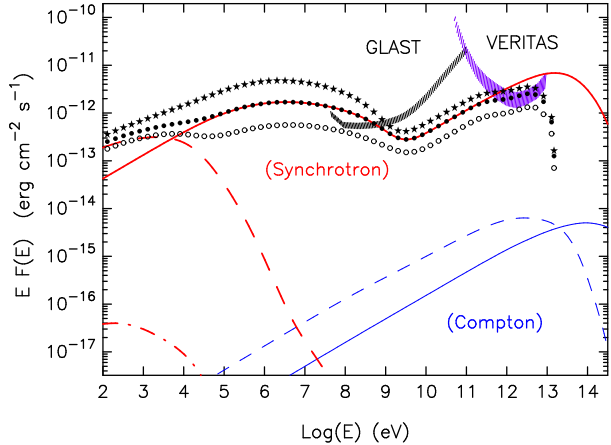}
\caption{Fluxes from the electromagnetic cascade initiated
within $0.5\,$Mpc distances from the core of Cyg A, assuming
an average magnetic field $B = 20~\mu$G in the cavity.
The total injection power of  UHECR protons
is $4\times 10^{46}\, \rm erg/s$. The photon fluxes expected for
$t_{age} = 30\,$Myr, 100\,Myr and 300\,Myr are shown by the open dots,
full dots and stars, respectively.
The solid, dashed, and dot-dashed curves show three generations of
synchrotron (heavy curves) and Compton (light curves) cascade radiations
in the case of $t=100\,\rm Myr$. Note that the third
generation of Compton cascade radiation is too weak to appear on the plot.
Sensitivities for the FGST and VERITAS are also shown.}
\vskip0.2in
\label{f2}
\end{figure}

Fig.\ \ref{f2} shows $\gamma$-ray fluxes calculated for the electromagnetic 
cascade initiated by the injection of $4\times 10^{46}\,\rm erg$ s$^{-1}$ 
of UHECR protons into the cavity, and further cosmic ray interactions with
 the EBL in the $r=0.5\,\rm Mpc$ region
surrounding Cyg A. Even though the EBL is dominated by the CMBR,  
interactions with the diffuse infrared/optical radiation  are also 
included in the calculations, using the EBL of \citet{pbs05}.
The injection spectrum of UHECR protons is a power-law with an $\alpha
= -2.1$ number spectral index with a low-energy cutoff at $E = 1$ TeV and a high-energy
exponential cutoff at $E = 3\times 10^{20}$
eV. We assume a mean magnetic field $B=20~\mu G$.
Escape of particles is given by the Bohm diffusion
approximation in a single zone model. The method
of calculation follows the approach described by \citet{ad03}.

The solid, dashed, and dot-dashed curves in Fig.\ \ref{f2} show the contributions to the
fluxes from the photohadronic secondary electrons and the first two generations of
cascade electrons for injection ages $t_{age}=  10^8 \,\rm yr$. 
The secondary electrons also include
the electrons from $\pi^0$-decay $\gamma$ rays that convert promptly into
electron-positron pairs inside the source. In the strong magnetic field of the
Cyg A environment, the Compton flux is dominated by synchrotron radiation
from the photohadronic secondaries and the first generation of cascade electrons.

The open dots,
full dots and stars in Fig.\ \ref{f2} show the received $E \cdot F(E)$ spectral energy fluxes
for injection ages
$t_{age} = 30\,$Myr, 100\,Myr and 300\,Myr,
respectively. The lower and higher energy peaks in the spectral energy distributions
 primarily result from
photopair and photomeson processes, respectively. The sensitivities for a
one-year, $5\sigma$ detection of a point source with
the FGST in the scanning mode,\footnote{www-glast.slac.stanford.edu/software/IS/glast\_lat\_performance.htm;
note that the FGST sensitivity will be somewhat poorer than shown
at the $5.76^\circ$ galactic latitude of Cyg A due to Galactic background.} and for a
50 hour,
5$\sigma$ detection with VERITAS are shown. The lower bound of the VERITAS sensitivity applies
to a point source, and the upper bound to a source of $25^\prime$ extent.

\section{Discussion and Conclusions}

If the radio lobes of Cyg A are powered by UHECR production
from the inner pc-scale jets, then trapping of these particles in the 
surrounding strong
magnetic-field region  leads to the production of secondary $\gamma$ rays that should be
significantly detected with the FGST in one year of observation if
the jet injection age is $\gtrsim 100$ Myr. If radio galaxies
are not the sources of UHECRs, then Cygnus A will not be detected 
by the FGST.
Cyg A could also be detected with VERITAS in a 50 hour pointing, depending 
on the duration of activity
of the central engine and the level of the EBL.

Detection of GeV $\gamma$ rays from Cyg A with the FGST might
also be expected to arise from other processes.
The $\sim 10$ -- 100 GeV radio-emitting electrons
from the lobes
of radio galaxies will Compton-scatter CMB photons to MeV -- GeV
energies \citep[e.g.,][]{che07}.  For the strong magnetic field, $\approx 60~\mu$G,
in the lobes of Cyg A, however,
the ratio of the magnetic-field to CMBR energy densities is
$\approx 400$. Thus the total energy flux of
Compton-scattered CMBR from Cyg A is $\approx 10^{45}$
erg s$^{-1}/[400(4\pi d^2)] \cong 4\times  10^{-13}$ ergs cm$^{-2}$
s$^{-1}$, with the $E\cdot F(E)$ flux a factor of $\sim 5$ -- 10 lower.
As can be seen from Fig.\ \ref{f2},
this process is almost two orders of magnitude below the UHECR-induced synchrotron flux,
and well below the FGST sensitivity.

{
\citet{ias05} considered fluxes expected from the $\lesssim 1 \,\rm Mpc$ halos of clusters of
galaxies with weaker magnetic fields, $B\simeq 0.1$-$1\,\mu$ G in a model
where acceleration of UHECRs occurs in accretion shocks in the cluster. Because of
lower maximum energies of accelerated protons, $E\lesssim 10^{19}\,\rm eV$, and
lower magnetic fields, this model predicts hard spectral fluxes of Compton origin
peaking at TeV energies}.
\citet{ga05} predicted that synchrotron radiation from
$\gtrsim 10^{18}\,\rm eV$ electrons is produced by
secondaries of UHECRs that leave the acceleration region
and travel nearly rectilinearly through
weak intergalactic magnetic fields at
the level $B \sim 10^{-7}$ -- $10^{-9}\,$G. These sources would appear
as point-like quiescent GeV -- TeV sources with spectra in the
GeV domain as hard as $\alpha \cong -1.5$, and very soft,
$\alpha \lesssim -3$ spectra in the 100 GeV -- TeV
domain.

{ In contrast to both these models}, we predict
soft 0.1 -- 1 GeV spectra with $\alpha \cong -2.5$ and hard, $\alpha \cong -2$
spectra at TeV energies due to the much
higher magnetic field in the confinement region. These models can
be clearly distinguished if Cygnus A is resolved by the Fermi Gamma ray Space 
Telescope or the ground-based
$\gamma$-ray telescopes, as the emission region in our model subtends an angle
$\approx 10^\prime$.

Because Cygnus A lies outside the GZK horizon, only UHECRs with energy
below the GZK energy could be correlated with this source. Other closer
FRII radio galaxies that are correlated with the arrival directions of
UHECRs are, however, candidate sources of $\gamma$ rays made through
the mechanism proposed here. IGR J21247+5058 at $z = 0.02$ or $d\approx 80$
Mpc, recently  discovered with INTEGRAL \citep{mol07}, is 2.1 degrees
away from a HiRes Stereo event with $E > 56$ EeV (C.\ C.\ Cheung, private
communication, 2008).\footnote{This radio galaxy was identified as such only
recently and would not have appeared in the list of AGN used by the HiRes
collaboration in their correlation study \citep{abb08}.} PKS 2158-380 at
$\approx 140$ Mpc is also within 3.2$^\circ$ degrees of an Auger UHECR with
$E> 57$  EeV \citep{mos08}. By comparison with Cyg A, these are low luminosity
FRIIs, and their predicted flux level will require detailed modeling for each
source, as done here for Cyg A.
Variability of $\gamma$-ray emission would rule out our model.


\acknowledgments

We thank Vladimir Vassiliev for discussion and providing the VERITAS sensitivities,
Teddy Cheung and Felix Aharonian for important comments, and the referee for a detailed
report.
The work of AA and visits to NRL were supported by the
GLAST Interdisciplinary Science Investigation Grant DPR-S-1563-Y.
The work of CDD is supported by the Office of Naval Research.


\begin{thebibliography}{}

\bibitem[Abbasi et al.(2008)]{abb08} Abbasi, R.~U., et al.\
2008, ArXiv e-prints, 804, arXiv:0804.0382

\bibitem[Aharonian et al.(1994)]{acv94}
Aharonian, F.~A., Coppi, P.~S., V\"olk, H.~J.\ 1994, ApJ, 423, L5

\bibitem[Aharonian(2002)]{aha02}
Aharonian, F.~A.\ 2002, MNRAS, 332, 202

\bibitem[Atoyan \& Aharonian(1999)]{aa99} Atoyan, A.~M., \&
Aharonian, F.~A.\ 1999, MNRAS, 302, 253

\bibitem[Atoyan \& Dermer(2001)]{ad01}
    Atoyan, A., and Dermer, C.\ 2001, PRL 87, 221102

\bibitem[Atoyan \& Dermer(2003)]{ad03}
    Atoyan, A., and Dermer, C.\ 2003, ApJ, 586, 79



\bibitem[Begelman
\& Cioffi(1989)]{bc89} Begelman, M.~C., \& Cioffi, D.~F.\ 1989, \apjl, 345, L21

\bibitem[Carilli et al.(1996)]{car96} Carilli, C., Perley, 
R., Bartel, N., \& Dreher, J.\ 1996, in Cygnus A -- Study of a Radio Galaxy, 
eds.\ by C.L. Carilli and D.E. Harris (Cambridge: Cambridge University Press), 76

\bibitem[Carilli et al.(1991)]{car91} Carilli, C.~L., Perley, 
R.~A., Dreher, J.~W., \& Leahy, J.~P.\ 1991, \apj, 383, 554 

\bibitem[Carilli \& Barthel(1996)]{cb96} Carilli, C.~L., \& Barthel, P.~D.\ 1996, \aapr, 7, 1

\bibitem[Cheung(2007)]{che07} Cheung, C.~C.\ 2007, The First
GLAST Symposium, S.\ Ritz, P.\ Michelson, \& C.\ Meegan, eds.\ (AIP: New York) 921, 325



\bibitem[ Dermer \& Atoyan(2004)]{da04}
Dermer, C.~D., \&  Atoyan, A.\ 2004, ApJ, 611, L9


\bibitem[Feretti et al.(1995)]{fer95}
    Feretti, L., Dallacasa, D., Giovannini, G., \& Tagliani, A.\ 1995, A\&A,
     302, 680

\bibitem[Ferrari et al.(2008)]{fer08} Ferrari, C., Govoni, 
F., Schindler, S., Bykov, A.~M., 
\& Rephaeli, Y.\ 2008, Space Science Reviews, 134, 93 

\bibitem[Gabici \& Aharonian(2005)]{ga05}
Gabici, S., \& Aharonian, F. A., 2005, Phys. Rev. Lett. 95, 251102

\bibitem[Halzen \& Hooper(2002)]{hh02}
Halzen, F., \& Hooper, D.\ 2002, Reports of Progress in Physics, 65, 1025

\bibitem[Harari et al.(2006)]{hmr06} Harari, D., Mollerach,
S., \& Roulet, E.\ 2006, J.\ Cosmology and Astroparticle Physics, 11, 12

\bibitem[Inoue et al.(2005)]{ias05}
Inoue, S., Aharonian, F.~A., and Sugiyama, N. 2005, \apjl, 628, L9

\bibitem[Kim et al.(1990)]{kim90}
      Kim, K.-T., Kronberg, P. P., Dewdney, P. E., \& Landecker, T. L.\ 1990,
       ApJ, 355, 29

\bibitem[Molina et al.(2007)]{mol07} Molina, M., et al.\
2007, \mnras, 382, 937

\bibitem[Moskalenko et al.(2008)]{mos08} Moskalenko, I.~V.,
Stawarz, L., Porter, T.~A.,
\& Cheung, C.~C.\ 2008, ArXiv e-prints, 805, arXiv:0805.1260

\bibitem[M\"ucke et al.(1999)]{muc99}
   M\"ucke, Rachen, J.P., A., Engel, R., Protheroe, R.J., and Stanev, T.
  1999, Pub.\ Astron.\ Soc.\ Australia, 16, 160

\bibitem[Perley et al.(1984)]{pdc84} Perley, R.~A., Dreher,
J.~W., \& Cowan, J.~J.\ 1984, \apjl, 285, L35


\bibitem[Pierre Auger Collaboration(2007)]{Auger07}
The Pierre Auger Collaboration, et al., 2007, Science, 318, 938

\bibitem[Primack et al.(2005)]{pbs05} Primack, J.~R.,
Bullock, J.~S.,  \& Somerville, R.~S.\ 2005, High Energy Gamma-Ray Astronomy, 745, 23

\bibitem[Scheuer(1974)]{sch74} Scheuer, P.~A.~G.\ 1974, 
\mnras, 166, 513

\bibitem[Smith et al.(2002)]{Smith02} Smith, D.~A., Wilson,
A.~S., Arnaud, K.~A., Terashima, Y., \& Young, A.~J.\ 2002, \apj, 565, 195



\bibitem[Steenbrugge \& Blundell(2008)]{sb08} Steenbrugge, K.~C., \& Blundell, K.~M.\ 2008, \mnras, 388, 1457 


\bibitem[Tadhunter et al.(2003)]{tud03} Tadhunter, C.,
Marconi, A., Axon, D., Wills, K., Robinson, T.~G.,
\& Jackson, N.\ 2003, \mnras, 342, 861

\bibitem[Wehrle et al.(1998)]{weh98}
    Wehrle, A.\ E.\ et al.\ 1998, ApJ, 497, 178


\bibitem[Wilson, Young, and Shopbell(2001)]{wys01}
    Wilson, A. S., Young, A. J., and Shopbell, P. L. 2001, ApJ, 547, 740

\bibitem[Wilson et al.(2006)]{wil06}
Wilson, A., et al. 2006, ApJ, 644, L9
\end{thebibliography}
\end{document}